\documentclass[3p,,preprint,12pt]{elsarticle}

\usepackage{tabulary,xcolor}
\usepackage{amsfonts,amsmath,amssymb}
\usepackage[T1]{fontenc}
\makeatletter
\let\save@ps@pprintTitle\ps@pprintTitle
\def\ps@pprintTitle{\save@ps@pprintTitle\gdef\@oddfoot{\footnotesize\itshape \null\hfill\today}}
\def\hlinewd#1{%
  \noalign{\ifnum0=`}\fi\hrule \@height #1%
  \futurelet\reserved@a\@xhline}

\AtBeginDocument{\ifNAT@numbers \biboptions{sort&compress}\fi}
\makeatother

\usepackage{ifluatex}
\ifluatex
\usepackage{fontspec}
\defaultfontfeatures{Ligatures=TeX}
\usepackage[]{unicode-math}
\unimathsetup{math-style=TeX}
\else 
\usepackage[utf8]{inputenc}
\fi 
\ifluatex\else\usepackage{stmaryrd}\fi

\usepackage{url,multirow,morefloats,floatflt,cancel,textcomp,tfrupee}
\usepackage{pifont}
\usepackage[nointegrals]{wasysym}
\makeatletter

\AtBeginDocument{
\expandafter\ifx\csname eqalign\endcsname\relax
\def\eqalign#1{\null\vcenter{\def\\{\cr}\openup\jot\m@th
  \ialign{\strut$\displaystyle{##}$\hfil&$\displaystyle{{}##}$\hfil
      \crcr#1\crcr}}\,}
\fi
}

\let\lt=<
\let\gt=>
\def\processVert{\ifmmode|\else\textbar\fi}

\@ifundefined{subparagraph}{
\def\subparagraph{\@startsection{paragraph}{5}{2\parindent}{0ex plus 0.1ex minus 0.1ex}%
{0ex}{\normalfont\small\itshape}}%
}{}

\newcommand\role[1]{\unskip}
\newcommand\aucollab[1]{\unskip}
  
\@ifundefined{tsGraphicsScaleX}{\gdef\tsGraphicsScaleX{1}}{}
\@ifundefined{tsGraphicsScaleY}{\gdef\tsGraphicsScaleY{.9}}{}
\def\checkGraphicsWidth{\ifdim\Gin@nat@width>\textwidth
	\tsGraphicsScaleX\textwidth\else\Gin@nat@width\fi}

\def\checkGraphicsHeight{\ifdim\Gin@nat@height>.9\textheight
	\tsGraphicsScaleY\textheight\else\Gin@nat@height\fi}

\def\fixFloatSize#1{\@ifundefined{processdelayedfloats}{\setbox0=\hbox{\includegraphics{#1}}\ifnum\wd0<\columnwidth\relax\renewenvironment{figure*}{\begin{figure}}{\end{figure}}\fi}{}}
\let\ts@includegraphics\includegraphics

\def\inlinegraphic[#1]#2{{\edef\@tempa{#1}\edef\baseline@shift{\ifx\@tempa\@empty0\else#1\fi}\edef\tempZ{\the\numexpr(\numexpr(\baseline@shift*\f@size/100))}\protect\raisebox{\tempZ pt}{\ts@includegraphics{#2}}}}

\AtBeginDocument{\def\includegraphics{\@ifnextchar[{\ts@includegraphics}{\ts@includegraphics[width=\checkGraphicsWidth,height=\checkGraphicsHeight,keepaspectratio]}}}

\def\URL#1#2{\@ifundefined{href}{#2}{\href{#1}{#2}}}

\def\UrlOrds{\do\*\do\-\do\~\do\'\do\"\do\-}%
\g@addto@macro{\UrlBreaks}{\UrlOrds}
\makeatother

\begin{document}

\begin{frontmatter}
	
\title{Cahn-Hilliard diffuse interface simulations of bubble-wall collisions}
  
\author[a,b]{Sohrab Towfighi}
\ead{sohrab@alumni.ubc.ca}
\author[a,c]{Hadi Mehrabian\corref{contrib-fda151dad5448237e05827f39bf2e7bb}}
\ead{hadim@mit.edu}\cortext[contrib-fda151dad5448237e05827f39bf2e7bb]{Corresponding author.}
    
\address[a]{Department of Chemical and Biological Engineering\unskip, 
    University of British Columbia\unskip, Vancouver\unskip, V6T 1Z3\unskip, BC\unskip, Canada}
    
\address[b]{Faculty of Medicine\unskip, 
    University of Toronto\unskip, Toronto\unskip, M5S 1A8\unskip, ON\unskip, Canada}
  	
\address[c]{Department of Chemical Engineering\unskip, 
    Massachusetts Institute of Technology\unskip, Cambridge\unskip, 02139\unskip, MA\unskip, USA}  

\begin{abstract}
The collision of a rising bubble with a superhydrophilic horizontal surface is studied numerically using the Cahn-Hilliard diffuse-interface method. For the studied systems, the Bond number, $Bo=\rho R^2 g/\sigma$, varies between 0.25 to 1.5, and the Galileo number, $Ga=\rho (R^3 g)^{1/2}/\mu$, changes between 8.5 to 50. We assume that the viscosity and density of the bubble are negligible compared to the surrounding medium. We show that our computations reproduce experimentally observed dynamics for the collision of a bubble with a solid surface. Furthermore, for the studied range of parameters, the bubble-wall collision is mainly in the inertial regime. Our simulations show that even in the absence of substantial viscous dissipation, the ratio of rebound-to-collision velocity, the so-called coefficient of restitution, is much smaller than one for the studied systems. More importantly, the coefficient of restitution best scales with the Froude number, $Fr=U_{t}/\sqrt{g R}$, the ratio of inertial to gravitational forces.

\end{abstract}
	
\end{frontmatter}
    
\section{Introduction}
\begin{figure*}[ht]
\centering
\includegraphics[scale=0.10]{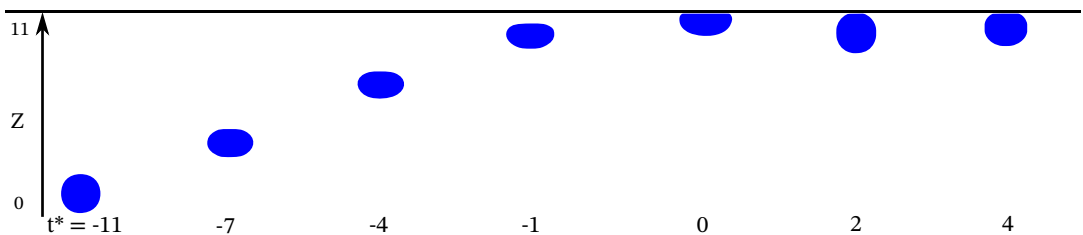}
\caption{Snapshots of the bubble rise for a system with $Bo = 0.5$, $Ga=40$ ($Re = 75$, and $We = 1.8$). Time and z-coordinate are in dimensionless form. 
}
\label{f:logo}
\end{figure*}

Bubbles hold great scientific interest because of their occurrence in many physical systems and their utility in numerous applications. The interaction between a bubble and a solid surface has received considerable attention experimentally \citep{Tsao1997b,Malysa2005,Hendrix2012}, computationally \citep{90Sho,10Omo,13Yue}, and analytically \citep{Brenner1961,Manica2015}. In this study, we are particularly interested in the impact of bubbles on a non-wetting horizontal surface. Consider a bubble inside a large liquid container, with a superhydrophilic surface on the ceiling of the tank as is illustrated in Fig.~\ref{f:logo}. The bubble-wall impact process for such a system can be best described in three stages: acceleration of the bubble from the initial static state until its velocity and shape become steady, deceleration on approaching the wall, and rebound. We aim to characterize the somewhat complicated collision process with a simple lump parameter called the coefficient of restitution, similar to what has been done for the characterization of the impact of solid particles and droplets with solid surfaces.

%
%
%
%

The initially static bubble accelerates under the action of buoyancy and reaches a steady shape and velocity after some time, the length of which depends on the balance of viscous and gravitational forces. For a highly viscous surrounding liquid, the Stokes flow regime occurs, meaning that the bubble remains spherical, and by equating the drag and buoyancy forces, its terminal velocity can be calculated. For a bubble moving at velocity $U_t$, if the inertial pressure, $\rho U_t^2/2$, dominates the Laplace pressure, $2 \sigma/R$, then surface tension will give way to inertia and the bubble will deform from its spherical shape. The ratio of these two pressures is called the Weber number $We=\rho U_t^2 R/\sigma$ and is an indication of the shape elongation in the direction normal to the flow, resulting in a non-spherical shape which will be called a spheroid shape hereafter. In most real cases, and also in our simulations, the $We$ number is larger than one, hence the bubble's shape deviates considerably from perfect sphericity. 
There are different regimes for the dynamics and shape of a rising bubble which are nicely delineated by Tripathi et al. \cite{Tripathi2015} in terms of the bubble Bond (E\"{o}tv\"{o}s) number, $Bo=\rho R^2 g/\sigma$, and Galileo number, $Ga=\rho (R^3 g)^{1/2}/\mu$.  At larger Bond number, another phenomenon that complicates bubble rise is the onset of path instability in the form of a zig-zagging or helical rise trajectory \citep{00Mag}. In this study, we limit ourselves to a regime with spherical or spheroidal bubbles without unsteady motion during the rise, which means that we restrict this study to the range of Bond and Galileo numbers which produce spheroidal or spherical bubbles without path instability.

The collision of a rigid particle with a solid wall can be characterized by a lumped parameter called the coefficient of restitution, defined as $\varepsilon=U_r/U_t$, in which $U_t$ and $U_r$ are the velocity of the particle before and after the collision, respectively. The coefficient of restitution is indicative of the dissipated energy during the impact. It is shown that the coefficient of restitution for the impact of solid particles with a rigid wall scales with Stokes number, which is the ratio of inertial to viscous effects \citep{Joseph2001,Gondret2002}. 
Similarly, Legendre et al. \cite{Legendre2005} found that the impact of the droplets on the solid wall could also be characterized by a modified Stokes number which includes the inertia of the associated surrounding fluid. It is interesting to examine the universality of this description for the collision of bubbles onto rigid surfaces. In fact, we will show that the bubble collision follows an entirely different scaling.

In reality, there are complications which bring uncertainty to the experimental results on bubble-wall collisions; the dynamics are usually affected by an array of subtle effects, including surfactants \citep{05Mal,10Zed}, wall roughness \citep{07Kra} and the presence of micro-bubbles near the hydrophobic wall which are unavoidable \citep{07Zaw}. This makes a computational study with complete control over input parameters necessary. Most intriguingly, Zenit and Legendre \citep{09Zen} suggested that the coefficient of restitution $\varepsilon$ can be fitted into two different expressions, either as a function of the modified Stokes number $St^*=C_m \rho R U_t/9\mu$ or as a function of a modified Ohnesorge number $Oh^*=\mu/\sqrt{\rho^* \sigma R}$, where $\sigma$ is the surface tension and $\rho^*=\rho_b + C_m \rho$ accounts for the added mass effect with the coefficient $C_m$ calculated from Lamb's classical formula, $C_m = ((\chi^2-1)^{1/2}-cos^{-1}\chi^{-1})/(cos^{-1}\chi^{-1}-(\chi^{2}-1)^{1/2}\chi^{-2})$. 
In this formula, $\rho_b$ is the density of bubble and $\chi$ is the bubble aspect ratio defined as $R_h/R_v$, in which $R_h$ and $R_v$ are the horizontal and vertical radius of the bubble spheroid. 
However due to uncertainty in their data and limited range of the input parameters they were unable to verify their hypothesis. Therefore, an investigation into the coefficient of restitution is warranted.

We use the Cahn-Hilliard diffuse interface method to simulate the collision of the bubble to a super-hydrophilic horizontal surface. It has been shown that this method is very successful in reproducing the experimental interfacial dynamics of a wide range of systems \citep{Mehrabian2011, Mehrabian2013, Mehrabian2014}. Similar to real interfaces, this model considers the finite thickness of the interface and captures its dynamics using an energy-based formalism whose details will be discussed in section \ref{s:method}.
This paper is organized as follows. We begin by detailing the problem setup and the computational method; then we demonstrate that the model can reproduce experimental work well. Subsequent the exposition, we dive into a presentation of the findings by giving a general description of collision phenomena, the scaling for bubble velocity and the coefficient of restitution, followed by a discussion. 

\section{Numerical simulations}

\subsection{Problem Setup}\label{s:setup}
The simulations are done in an axisymmetric geometry where a spherical bubble with radius $R$ is released with zero initial velocity at the bottom of a liquid cylindrical column. The bubble then rises along the axis of symmetry, whose height is 11$R$. This is illustrated in Fig.~\ref{f:logo}. The bubble is initially placed with its centroid at 1.5$R$ above the bottom boundary and 4$R$ from the side boundaries. Inspired by the experimental setup used by Zenit and Legendre \citep{09Zen}, we choose the computational domain size to be sufficiently large so that the side and bottom boundaries do not influence the bubble dynamics, and we verify this through numerical tests. This is facilitated by imposing a ``relaxed" boundary condition on the side and bottom boundaries: vanishing tangential velocity and normal stress equal to the hydrostatic pressure \citep{00Gre}. This allows flow through these boundaries and minimizes their confinement effect. 
The top wall has a no-slip boundary condition, and its contact angle is set to 180 degrees to make it completely non-wetting. 

The physical problem is determined by the following dimensional parameters: the bubble radius $R$, surface tension $\sigma$, the gravitational acceleration $g$, and the density and viscosity of the bubble and liquid phases denoted by $\rho_b$, $\mu_b$ and $\rho$, $\mu$, respectively. 
Besides the density ratio $\rho_{b}/\rho$ and viscosity ratio $\mu_{b}/\mu$, this system can be described by two dimensionless groups; we use the two groups suggested by Tripathi et al. \citep{Tripathi2015} to categorize the bubble dynamics, namely the Bond number $Bo=\rho g R^{2}/\sigma$ and the Galileo number $Ga=\rho (R^3 g)^{1/2}/\mu$. The bond number shows the ratio of gravitational to surface force, and Galileo number indicates the ratio of gravity to viscous force and is an indicator of the bubble terminal velocity. 


\subsection{Computational method}\label{s:method}
The computational method is based on the diffuse-interface framework \citep{04Yue,06Yue,07Yue} used to dynamically capture the location of the interface. To demarcate the interface between the bubble and the surrounding liquid, a phase-field parameter $\phi$ evolves as governed by the Cahn-Hilliard equation. This is solved together with the Navier-Stokes equation using a finite element solver on an unstructured triangular grid with the adaptive mesh refinement at the moving interface.
The numerical scheme uses fully implicit time-stepping and Newton's method to handle the nonlinear equations. Details of the theoretical model and numerical method and parameters can be found in previous publications \citep{04Yue,06Yue,07Yue}. 
In particular, the diffuse-interface model introduces two additional parameters to incorporate the nanoscale physics into macroscale simulations; the Cahn number $Cn=\epsilon/d$ which is an indicator of the interfacial thickness $\epsilon$, and the diffusion length scale $S=l_d/R$ which represents the Cahn-Hilliard diffusivity $l_d$. More detailed description of the parameter definitions can be found in \citep{Amphi3D}. For the diffuse-interface model to make accurate predictions, the interface must be sufficiently thin so that the sharp-interface limit is achieved. This indicates that the results are independent of the interfacial thickness and requires the interface to be adequately resolved by having locally refined grids \citep{Amphi3D}. Extensive testing of $Cn$ and $S$ values have been carried out before and the results validated against sharp-interface benchmarks \citep{06Yue,07Yue,Amphi3D}.
Our numerical experiments have shown that for $Cn < 10^{-2}$, the sharp-interface limit is approached. In the results to be presented here, we have used $ Cn = 10^{-2}$ when the bubble is far from the wall and we reduce to $Cn=6.0\times 10^{-3}$ when the bubble is close to the wall. This balances the high computational cost with the need for a thinner diffusion interface to accurately capture the dynamics of the thin layer between the bubble and the wall. 

\begin{figure*}
\centering
\includegraphics[scale=1]{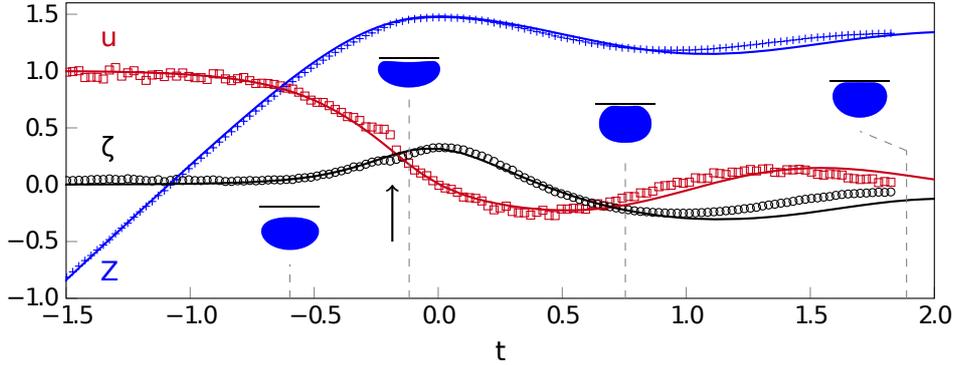}
\caption{Comparison of simulation results with experimental data of Ref. \citep{12Zen}. Experimental data is for a non-polar liquid, free from surfactant effects. Experimental values are given as points while simulated values are presented as curves; ({ $+$}), ({ $\square$}), and ($\circ$) represent the center of mass position $Z$, velocity $u$, and deformation $\zeta$, respectively. The dimensionless parameters for this case are $Ga = 26$, and $Bo=0.66$. The insets show axisymmetric bubble geometry at $t =$ -1.20, -0.24, 1.50, and 3.80. The arrow points to a point of discrepancy discussed in the text.}
\label{f:validation}
\end{figure*}

The choice of $S$ is a subtle issue \citep{07Yue,Yue2010} and its value should be calibrated by an experimental or theoretical data point. We found that $S=5\times 10^{-4}$ produces excellent agreement with the experiments as is shown in Fig.~\ref{f:validation}). Although the liquid in principle never dewets the upper wall, Cahn-Hilliard diffusion may exert a minor effect on the interfacial dynamics when the liquid film becomes very thin.

%
%
%
%

\subsection{Validation, and comparison to experiment}

In this section, we matched the dimensionless parameters used in our model with those from experiments for the rectilinear rise of an ultra-clean bubble and its collision with a glass surface. In figure \ref{f:validation}, we compared the numerically calculated values for the center of mass position, center of mass velocity, and the bubble's aspect ratio to the experimental data of Ref. \citep{12Zen} to validate the computational results. The experimental result is for a bubble with a clean interface; by using a non-polar oil, Zenit and Legendre were able to avoid complications due to surfactants \citep{12Zen}. The Galileo number and Bond number are $Ga = 26$, and $Bo=0.66$ for this system, respectively. The viscosity ratio matches the experimental value of $\mu_b/\mu=0.004$, although numerical experiments show the bubble viscosity to be negligible for $\mu_b/\mu<0.0055$. An exception to exact parameter matching is the density ratio. The experimental value for air-water density ratio, $\rho_b/\rho=0.001$, would be too small for accurate resolution in a diffuse-interface formalism \citep{06Yue}. In numerical experiments, we found that the bubble inertia becomes entirely negligible for $\rho_b/\rho < 0.01$, and have used $\rho_b/\rho =0.005$ for the computations. 
First, we computed the terminal velocity and shape of the bubble using different domain sizes and confirmed that the bubble dynamics are free from boundary effects. Then, we recorded the center of mass position for the bubble, $z_c$, as measured from the bottom of the geometry, the velocity of the bubble $u=dz_c/dt$, and the bubble aspect ratio $\chi=R_h/R_v$ as functions of time. All lengths are scaled by $R$ and velocities by $U_t$. To determine the aspect ratio, we measured the distance between the bubble's horizontal extremes (the horizontal radius $R_h$) and then applied conservation of volume to obtain the corresponding spheroid's vertical radius $R_v$.

Figure~\ref{f:validation} shows the simulation overlaying experimental data, with the bubble's center of mass position, instantaneous velocity, and aspect ratio plotted as functions of time. The origin of time is set at the point of zero velocity when the bubble centroid is roughly closest to the wall and the bubble at maximum deformation. Consistent with the data presentation in Ref. \citep{09Zen}, the bubble position is indicated by $Z=z_c-3.5$ and its deformation by $\zeta=\chi/\chi_0 - 1$, $\chi_0$ being the aspect ratio of the bubble.
Figure~\ref{f:validation} shows near perfect agreement between simulation and experiment. Note that the numerical solution closely tracks the experiment until the end of the first cycle. In later cycles, the discrepancy grows, with the numerical oscillation decaying more slowly than the experiment. We suspect that the deviation stems from the extended close interaction between the bubble surface and the wall, which in our model may have introduced extra friction due to interfacial diffusion. 
Also, we compared the computed terminal velocity and shape of the bubble between numerical and experimental results \citep{98Tom,08Tsa,13Yue}. The calculated bubble aspect ratio differs from experiments by $2.4\%$ to $7.1\%$ and the terminal velocity by 0.3\% to 7.4\%. These values are well within the scatter of experimental data, and the agreement is better than previous numerical results \citep{13Yue,08Tsa}.

\section{Results and discussion}\label{s:RD}

\subsection{Collision stages: acceleration, approach and rebound}

We can divide the bubble-wall collision process into three stages, namely, the acceleration, approach, and rebound stages \citep{09Zen}. In our simulations, the bubble starts from a static state, shown at $t=-11$ in Fig.~\ref{f:logo}, and accelerates due to the buoyancy force to its terminal velocity, $U_t$, and shape, $\chi_t$, at $t=-3$. The acceleration stage is not the focus of this study, so it is not shown in Fig.~\ref{f:validation}. 
The second stage, the approach, occurs as the bubble feels the presence of the wall which happens between $t=-3$ and $t=0$ in Fig.~\ref{f:validation}. During this stage, the bubble velocity decreases while its deformation increases; some part of the incoming kinetic and gravitational energy transforms into the surface energy, and part of it gets dissipated by the viscous forces. As the bubble approaches the wall, its deceleration is accompanied by drainage of the liquid film between the bubble and the wall. This leads to the formation of a central dimple and the appearance of the rim at its closest point to the wall. Eventually, bubble velocity vanishes at $t=0$, defined as the point of collision. However, as we will discuss later, this does not mean that the flow field around the bubble vanishes at this stage. 

The third stage, called rebound, starts when the bubble reaches the zero velocity point at $t=0$. During the rebound, the bubble's centroid velocity increases to a maximum value in the downward direction and then returns to zero at the end of the rebound at $t=1$. At the end of rebound, the bubble's center of mass is at its farthest position from the wall. For an entirely elastic collision, the bubble would be far enough from the wall to have a full cycle of acceleration, approach, and rebound again. However, in reality, there is energy loss, and another cycle of acceleration, approach and rebound could only occur if the bubble gets far enough from the wall at the end of its initial rebound. The bubble can undergo several rounds of oscillation, each time marked by an exchange between surface energy, gravitational energy, and kinetic energy while viscous forces dissipate the energy of the system. In the next section, we will discuss that viscous dissipation is not the only source of the momentum loss for studied bubble-wall collisions.
Fig.~\ref{f:validation} shows the second round of oscillation. In the end, at $t \approx 2$ in  figure~\ref{f:validation}, all the kinetic energy has been used, and the bubble has come to rest against the solid top wall.

\subsection{Scaling for the coefficient of restitution}\label{s:eps}

%
\begin{figure}
\centering
\includegraphics[scale=.3]{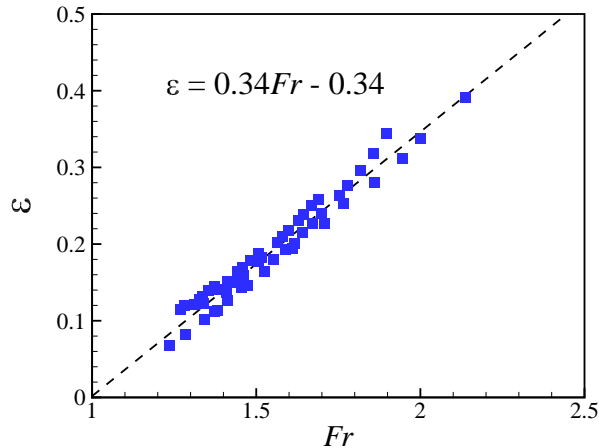}
\caption{Coefficient of restitution as a function of the Froude number.} 
\label{f:eps}
\end{figure}
%
An important parameter that characterizes the collision of solid particles or droplets is the ratio between their momentum before and after the collision, the so-called coefficient of restitution, $\varepsilon$. This parameter is a measure of the dissipated energy during the collision. Analogously, it is interesting to measure this quantity for the bubble-wall collision. The most natural way to define this ratio for the bubble-wall collision is to use the bubble's terminal velocity to characterize the incoming momentum and the maximum velocity of the bubble during its rebound ($U_r$ in Fig.~\ref{f:validation}) to determine the return momentum.

\begin{equation}\label{e:eps}
\varepsilon = - U_{r} / U_{t}, 
\end{equation}

The negative sign accounts for the change in direction during collision. It has been shown that the coefficient of restitution for the droplet or particle collisions scales with the ratio of the inertial to viscous forces, which can be measured either by the Reynolds or Stokes number. However, we found that for the bubble-wall collision the scaling is different. We tested all governing dimensionless groups and discovered that the best collapse of our numerical data towards a single parameter linear expression happens when $\varepsilon$ is plotted against the Froude number, defined as $Fr = U_{t}/\sqrt{g R}$. All the data from our simulations for the coefficient of restitution is presented in Fig.~\ref{f:eps} which show that the ratio of the maximum rebound velocity to terminal velocity increases almost linearly with the ratio of inertial to gravitational force, i.e., the $Fr$ number. The Froude number manifesting itself as the most dominant control parameter needs further investigation, as this scaling differs from that of the collision of droplets and particles. Moreover, if we consider the magnitude of the coefficient of restitution in Fig.~\ref{f:eps}, we see that they are less than 0.4, suggesting that there is a considerable momentum loss during the collision process. However, both gravity and inertia, whose ratio creates the Froude number, are conservative forms of energy, and it seems contradictory to represent the momentum loss in terms of conservative forces. In other words, if inertia and gravity govern the collision, then why is there a considerable reduction in the bubble momentum after the collision?
\begin{figure*}
\centering
\includegraphics[scale=1]{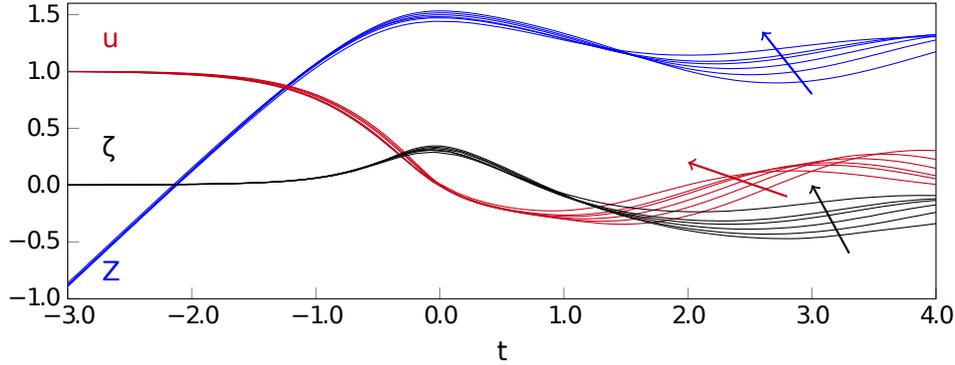}
\caption{Bubble approach is mainly inertial. For the shown systems, $Bo = 0.5$ and the corresponding $Ga$ changes from 20 to 50.} 
\label{f:selfsim}
\end{figure*} 
%
%
Commonly, when there is an energy loss in the fluid system, the first culprit is the viscous dissipation. The critical parameter that represents the relative significance of the viscous effects is the $Re$ number of the bubble. In our simulations, the minimum Reynolds number is 10 which shows the relative dominance of the inertial effects. Also, an interesting observation from our simulations is that the timescale of the approach stage mainly correlates with the inertial timescale. In other words, if we scale the velocity with the terminal velocity of the bubble and the time with the characteristic inertial time scale defined as $t_c=\frac{R}{U_t}$ then all velocity profiles collapse almost onto a single curve. In figure~\ref{f:selfsim}, the dimensionless velocity profile, $U(t)/U_t$, the position of bubble's center of mass, and the aspect ratio are plotted for cases with varying viscosity of surrounding liquid. These results suggest that the approach dynamics is independent of the viscosity of the liquid due to negligible viscous force. This is similar to the inertial drainage of the interstitial fluid for droplet and solid particle collision with a rigid wall. If viscous dissipation is not considerable, then why is there a significant decrease in the momentum of the bubble after the collision?

\begin{figure*}
\centering
\includegraphics[scale=0.4]{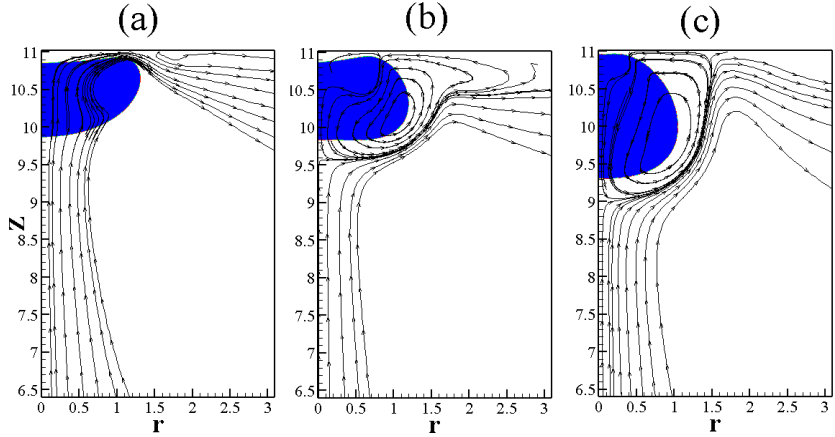}
\caption{Flow field around the bubble during the rebound for $Bo=0.5$ and $Ga=40$. (a) Flow field around the bubble at the end of the approach (beginning of the rebound). (b) and (c) Flow field around the bubble during the rebound.}
\label{f:under}
\end{figure*}

We believe that the answer is rooted in the importance of the added mass effect for bubble dynamics. 
Due to the negligible density of the bubble, the kinetic energy is stored mainly in the fluid surrounding the bubble. Since the surrounding fluid is not governed by surface tension, not all of the incoming kinetic energy will be transformed into surface energy during the approach. Some part of the kinetic energy will be changed into radial momentum (the direction parallel to the top wall in Fig.~\ref{f:logo}) and will not be recovered during the rebound to push the bubble the bubble away from the wall. It is important to note that the bubble rebound starts while the surrounding liquid is still moving toward the wall in the z-direction.
This is illustrated in a series of snapshots in Fig.~\ref{f:under}. In Fig.~\ref{f:under}(a), the bubble reaches its maximum spreading and starts to retract, while the velocity field in the surrounding fluid is in the upward direction. Such a velocity field will hamper the bubble rebound and creates another dimple on the lower surface of the bubble. Therefore, even in the absence of the considerable viscous dissipation, the bubble-wall collision could result in a major momentum loss. 

It is interesting to formulate any problem in terms of the dimensionless input parameters. If we fix the surface properties of the wall, the density ratio, and the viscosity ratio, then, of the five-dimensional groups, only two dimensionless groups remain independent. Therefore, any quantities of interest in our problem, such as $\varepsilon$ should be determined by two dimensionless groups. We showed that $Fr$ number is the most important parameter governing the bubble-wall collision for the cases considered in this study. However, we would like to obtain more accurate scaling by considering other dimensionless groups, which include viscous effects and accurately account for the inertial forces close to the wall, in order to give a more precise description of the collision.
We found it a daunting task, and none of the famous dimensionless parameters improve the scaling. We believe that the difficulty arises due to inaccurate representation of the viscous forces and the complicated contribution of the added mass effect during the collision and rebound. In other words, none of the typical dimensionless parameters can accurately represent the inertial and viscous effects.

%
%
%
\subsubsection{Comparison to other data on coefficient of restitution}
\begin{figure}
\centering
\includegraphics[scale=0.3]{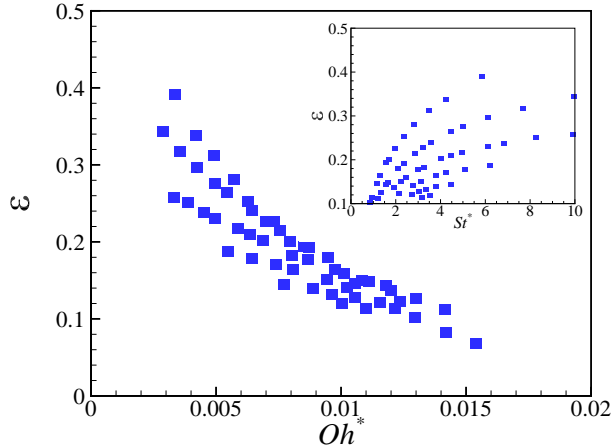}
\caption{Coefficient of restitution is plotted versus the previously suggested scaling parameters, i.e., the modified Ohnesorge number and the modified Stokes number which does not show any clear trend similar to the $Fr$ number scaling.}
\label{f:eps-other}
\end{figure}

The only previous work on the coefficient of restitution for bubble-wall collisions was done by Zenit and Legendre \cite{09Zen}, in which it is suggested that the coefficient of restitution might scale with the modified Stokes number, $St^*$, or the modified Ohnesorge number, $Oh^{*}=\sqrt{Ca/St^{*}}$. Therein it is stated that "In particular, it would be important to obtain measurements (through experiments or numerical simulations) for systems in which $Ca$ and $St^*$ could be varied independently. Moreover, it is necessary to perform experiments considering clean fluids to obtain a precise measurement of the critical value of the Stokes number beyond which a rebound can be expected." We tested the suggested scaling by plotting our numerical data versus these parameters, $St^{*}$ and $Oh^{*}$, in Fig.~\ref{f:eps-other}(a) and (b), respectively, which shows a considerable scatter with respect to either of them. Therefore, our numerical simulations revealed that such scaling does not hold for the large range of parameters in an entirely clean system.
Zenit et al. \citep{09Zen} mentioned that the presence of surfactants in the water might cause uncertainty in the accurate measurement of bubble dynamics close to the wall. It has been observed that presence of any contamination can dramatically affect the dynamics of the bubble during the collision \citep{Manica2016}. Accurate modelling in the presence of surfactants necessitates inclusion of a history force and modification of boundary conditions in the lubrication layer between the bubble and wall. Also, the terminal velocity for bubbles in contaminated systems is considerably lower than the clean systems \citep{Manica2016}. 
This makes the computational data much more valuable in the sense that we have control over the parameters and the measured dimensionless numbers are correct. The computational code and the numerical method have been successfully applied to the dynamics of the fluid interface in numerous settings. The only probable issue within the framework of the diffuse interface method is that interface and wall interact due to their diffusive nature of interface model, which may produce additional friction during the bubble spreading and retraction. We verified that the numerical simulations could correctly reproduce the experimental results for the clean system (Fig.~\ref{f:validation}) but caution is appropriate with respect to the role of diffusion in modeling the bubble-wall collision within the diffuse interface framework. 

%
%
%
%
\section{Conclusions}
We used the Cahn-Hilliard diffuse interface method to study the momentum loss of bubbles during collision with a rigid superhydrophilic horizontal wall. We simulated the bubble starting from the stasis and ensured that it reached a steady state before reaching the wall. The bubble-wall collision happens in two stages: the approach and rebound stages. For the studied systems (Bond numbers between 0.25 to 1.5, Galileo numbers between 8.5 to 50, and $Re$ numbers from 10 to 100) deceleration of the bubble before its collision to the wall is governed by the inertial forces due to relatively weak viscous effects. 
The interesting observation is that even in the absence of strong viscous forces, there is a considerable momentum loss after the impact due to the fact the kinetic energy of the surrounding liquid does not get completely converted to the surface energy in the approach stage. Furthermore, the coefficient of restitution for the studied systems best scales with the Froude number suggesting that gravitational deceleration is the primary deceleration mechanism during the rebound stage.

\section{Acknowledgments}
This research was partially supported by the NSERC, the Canada Research Chair program and the Canada Foundation for Innovation. We thank Roberto Zenit for generously providing us experimental data. We are grateful for discussions with James J. Feng, who played a substantial role in forming the work.

\section{References}
\bibliographystyle{jfm}


\end{document}